\pdfoutput=1
\documentclass[12pt]{article}\usepackage{quibbs}
\begin{document}

\title{Quibbs, a Code Generator\\
for Quantum Gibbs Sampling}

\author{Robert R. Tucci\\
        P.O. Box 226\\
        Bedford,  MA   01730\\
        tucci@ar-tiste.com}

\date{ \today}

\maketitle

\vskip2cm
\section*{Abstract}

This paper introduces
Quibbs v1.3,
a Java application
available for free. (Source code
included in the distribution.)
Quibbs is a ``code generator" for
quantum Gibbs sampling:
after the user inputs some files that specify
a classical Bayesian network,
Quibbs outputs
a quantum circuit
for  performing Gibbs sampling of that
Bayesian network
on a quantum computer. Quibbs
implements an algorithm
described in earlier papers,
that combines
various apple pie techniques such as:
an adaptive fixed-point version of
Grover's algorithm, Szegedy operators,
quantum phase estimation
and quantum multiplexors.

\section{Introduction}

We say a unitary operator
acting an array of qubits has been
compiled if
it has been expressed
as a Sequence of Elementary
Operations (SEO), where by elementary
operations we mean
 1 and 2-qubit operations such
as CNOTs and single-qubit rotations.
SEO's are often represented as quantum circuits.

There exist software, ``general quantum compilers"
(like Qubiter, discussed in Ref.\cite{TucQubiter}),
for compiling arbitrary unitary
operators (operators that have no a priori
known structure).
There also exists
software,``special purpose
quantum compilers"(like
each of the 7
applications
of QuanSuite,
discussed in Refs.\cite{quantree,quanfou,quanfruit}),
for
compiling unitary operators
that have a very definite, special
structure which is known a priori.

This paper introduces\footnote{The reason
for releasing the first public
version of Quibbs with such an odd
version number is that Quibbs shares
many Java classes with other previous Java
applications of mine
(QuanSuite
discussed in Refs.\cite{quantree,quanfou,quanfruit},
QuSAnn
discussed in Ref.\cite{TucQusann},
and Multiplexor Expander
discussed in Ref.\cite{TucQusann}),
so I have made the
decision to give all these applications
a single unified version
number.}
Quibbs v1.3,
a Java application
available for free. (Source code
included in the distribution.)
Quibbs is a ``code generator" for
quantum Gibbs sampling:
after the user inputs some files that specify
a classical Bayesian network,
Quibbs outputs
a quantum circuit
for  performing Gibbs sampling of that
Bayesian network
on a quantum computer.
Quibbs is not really a quantum
compiler (neither general nor special)
because, although it
generates a quantum circuit like
the quantum compilers do,
it doesn't
start with an explicitly
stated unitary matrix
as input.

Quibbs
implements the algorithm
of Tucci discussed in Refs.
\cite{TucGibbs,TucZ, TucGrover}.
The quantum circuit
generated by Quibbs
includes some quantum multiplexors.
The Java application Multiplexor Expander
(see Ref.\cite{TucQusann})
allows the user to replace
each of those multiplexors by
a sequence of more elementary
gates such as multiply
controlled NOTs and qubit rotations.
Multiplexor Expander is also
available for free, including source code.

For an explanation of the mathematical
notation
used in this paper, see some of my previous papers;
for instance, Ref.\cite{notationNAND} Section 2.

Throughout this paper, we will
often refer to an operator $V$.
$V$ is defined by
figure 5 of
Ref.\cite{TucGibbs}.
We will also use the acronym
AFGA (Adaptive Fixed-point Grover Algorithm)
for the algorithm described in Ref.\cite{TucGrover}.

\section{The Control Panel}
\label{sec-control-panel}

\begin{figure}[h]
    \begin{center}
    \includegraphics[scale=.70]{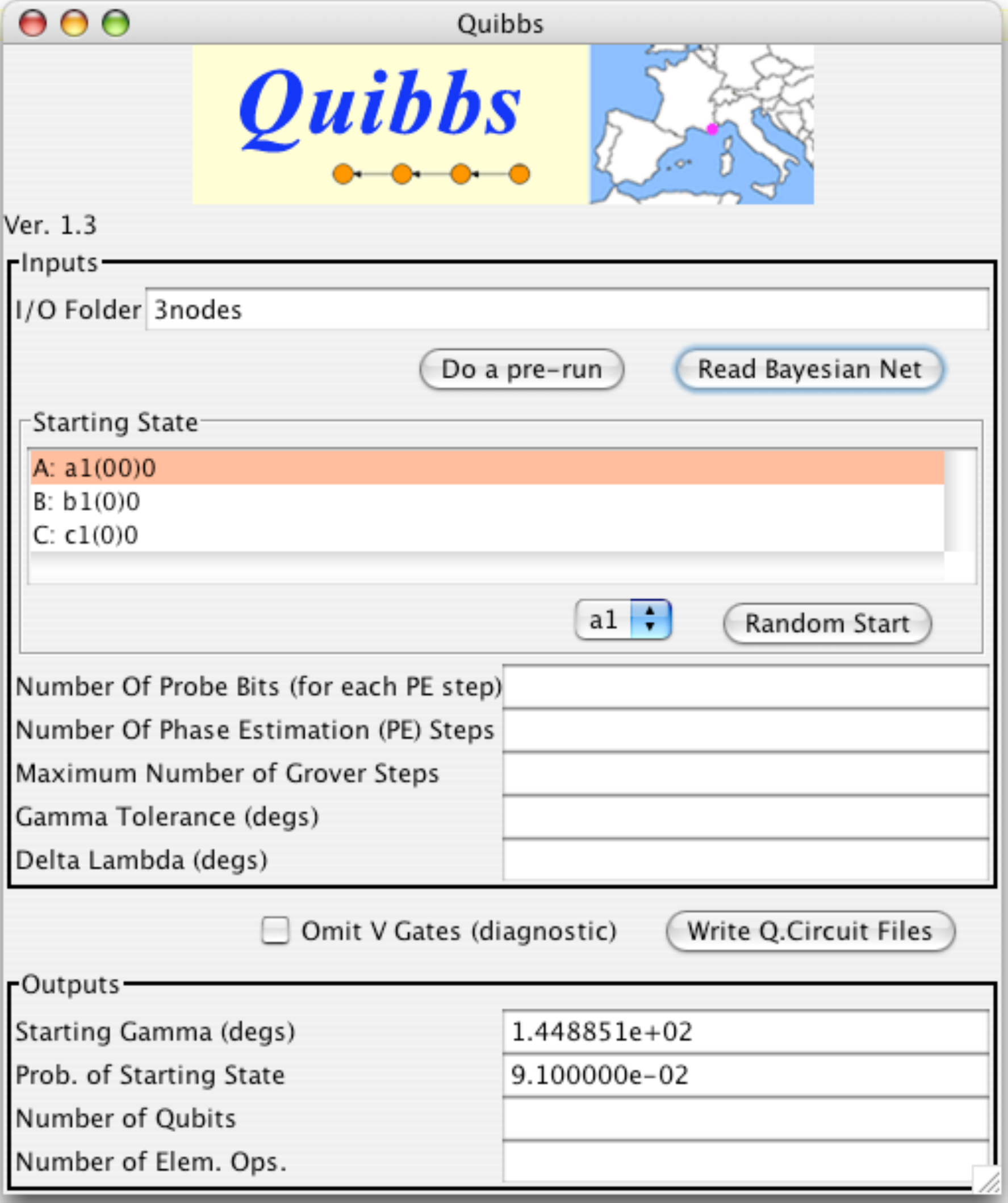}
    \caption{{\bf Control Panel} of Quibbs}
    \label{fig-quibbs-main}
    \end{center}
\end{figure}

Fig.\ref{fig-quibbs-main} shows the
{\bf Control Panel} for Quibbs. This is the
main and only window of Quibbs
(except for the occasional
error message window). This
window is
open if and only if Quibbs is running.
\newpage

The {\bf Control Panel}
allows you to enter the following inputs:
\begin{description}

\item[I/O Folder:] Enter in this
 text box the name
of a folder. The folder will contain Quibbs' input and
output files for the particular
Bayesian network
that  you are currently considering.
The I/O folder must
be in the same directory as
the Quibbs application.

To generate
a quantum circuit,
the I/O folder must contain
the following 3 input files:
\begin{itemize}
\item[(In1)] {\tt parents.txt}
\item[(In2)] {\tt states.txt}
\item[(In3)] {\tt probs.txt}
\end{itemize}
A detailed description of these 3 input files
will be given
in the next section. For this section,
all you need to
know is that: The {\tt parents.txt} file
lists the parent nodes of
each node of the Bayesian net being considered.
The {\tt states.txt} file lists the names of
the states of each node of the Bayesian net.
And the {\tt probs.txt} file gives the probability
matrix for each node of the Bayesian net.
Together, the In1, In2 and In3 files fully
specify the Bayesian network being
considered.

In Fig.\ref{fig-quibbs-main},
``3nodes" is entered in the {\bf I/O Folder}
text box.
A folder called ``3nodes" comes
with the distribution of Quibbs.
It contains, among other
things,  In1, In2, In3 files
that specify one possible Bayesian network
with 3 nodes. The Quibbs distribution
also comes with 3 other examples of
I/O folders. These are named
``2nodes", ``4nodeFullyConnected" and
``Asia".

When you press the
{\bf Read Bayesian Net}
button, Quibbs reads
files In1, In2 and In3.
The program then creates data structures
 that contain
complete information about
the Bayesian network. Furthermore, Quibbs fills the
scrollable list in the {\bf Starting State} grouping
with information that specifies
``the starting state". The starting state is
one particular instantiation (i.e.,
a particular state for each node) of the
Bayesian network.
Each row of the scrollable list
names a different node, and a particular
state of that node.
 For example,
Fig.\ref{fig-quibbs-main} shows the
Quibbs {\bf Control Panel} immediately after
pressing the {\bf Read Bayesian Net} button.
In this example, the Bayesian net
read in has 3 nodes called
$A, B$ and $C$, and the starting state has node
$A$ in state $a1$, node $B$ in state $b1$
and node $C$ in state $c1$.

Suppose $A$ is a node of the
Bayesian net being considered. And suppose
$A$ has $N_A$ states. Quibbs will
give each state of node $A$
a ``decimal name"; that is, a
number from 0 through $N_A-1$.
The ``binary name" of a state
is the binary representation
of its decimal name.
As shown in Fig.\ref{fig-quibbs-main},
the scrollable list
of the {\bf Control Panel}
gives not only the ``english name"
of the state of each node, but
also
the binary and decimal names of that state.
For example,
 Fig.\ref{fig-quibbs-main}
 informs us that state $a1$ of node $A$
 has binary name $(00)$ and decimal name $0$.

If you press the {\bf Random Start} button,
the starting state
inside the scrollable list
is changed to a randomly
generated one. Alternative, you can
choose a specific state for each node of
the Bayesian net by using the {\bf Node State Menu},
the menu immediately to the left of
the {\bf Random Start} button.
To use the {\bf Node State Menu},
you select the particular row of
the scrollable list that you want
to change. The {\bf Node State Menu}
mutates to reflect your row selection
in the scrollable list. You
can choose from the menu a particular node state.
When you do so, the selected row
in the scrollable list changes to reflect
your menu choice.

When you press the
{\bf Do a Pre-run} button,
Quibbs both reads and writes files:
it reads files In1 and In2 (but not In3,
so if In3 is not included in the I/O folder,
this button still works), and
it writes the following files, whose
content will be described later:

\begin{itemize}
\item {\tt probsF.txt}
\item {\tt probsT.txt}
\item {\tt blankets.txt}
\item {\tt nits.txt}
\end{itemize}

\item[Number of Probe Bits (for each PE step):]
This is the parameter $a=1,2,3,\ldots$
 for the operator $V$.

\item[Number of Phase Estimation (PE) Steps:]
This is the parameter $c=1,2,3,\ldots$
for the operator $V$.

\item[Maximum Number of Grover Steps:]
Quibbs will stop iterating the AFGA if
it reaches this number
of iterations.

\item[Gamma Tolerance (degs):]
This is an angle
given in degrees.
Quibbs will stop iterating the AFGA
if the absolute value of
$\gamma_j$ becomes smaller than this tolerance.
($\gamma_j$
is an angle in AFGA that tends to zero
as the iteration index $j$ tends to infinity.
$\gamma_j$ quantifies
how close the AFGA is to reaching the
target state).

\item[Delta Lambda (degs):]
This is the angle $\Delta\lambda$ of
AFGA,
given in degrees.
\end{description}

Once Quibbs has
successfully read files In1, In2 and In3,
and once you have filled
all the text boxes in the {\bf Inputs}
grouping, you can
successfully press the
{\bf Write Q. Circuit Files} button.
This will cause Quibbs to write
the following output files within
the I/O folder:

\begin{itemize}
\item[(Out1)] {\tt quibbs\_log.txt}
\item[(Out2)] {\tt quibbs\_eng.txt}
\item[(Out3)] {\tt quibbs\_pic.txt}
\end{itemize}
The contents of these
 3 output files will be described
in detail in the next section. For this
section, all you need to know
is that: The {\tt quibbs\_log.txt} file
records all the input and output
parameters that you entered into the
{\bf Control Panel}, so
you won't forget them.
The {\tt quibbs\_eng.txt} file
is an in``english" description
of a quantum circuit.
And the {\tt quibbs\_pic.txt} file
translates, line for line,
the english description found in
{\tt quibbs\_eng.txt}
into a ``pictorial" description.

Normally, you want to
press the
{\bf Write Q. Circuit Files} button
without check-marking the
{\bf Omit V Gates (diagnostic)} check box.
If you do check-mark it,
you will still generate
files Out1,Out2, Out3,
except that those files will
omit to mention all those gates
that generate the operator $V$,
at every place were it would
normally appear. Viewing the circuit
without its $V$'s is useful for
diagnostic and educational purposes,
but such a circuit is of course
useless for Gibbs sampling
the Bayesian net being considered.

The {\bf Control Panel}
displays the following
output text boxes.
(The {\bf Starting Gamma (degs)}
output text box
and the {\bf Prob. of Starting State}
output text box
are both filled as soon as
a starting state is given in
the inputs. The other output text boxes
are
filled when you press the
{\bf Write Q. Circuit Files} button.)

\begin{description}
\item[Starting Gamma (degs):]
This is $\gamma_0$, the first $\gamma_j$
in AFGA, given in degrees. In the
notations of Ref.\cite{TucGibbs},
and \cite{TucGrover},

\beqa
\gamma_0 &=& \acos(\hat{s'}\cdot\hat{t})=
2 \;\acos(|\av{s'|t}|)\\
&=& 2\;\acos(\sqrt{P(x_0)})
\;,
\eeqa
where $P(x_0)$ is the {\bf Prob. of
Starting State} defined next.

\item[Prob. of Starting State:]
This is the probability
$P(x_0)$ in Ref.\cite{TucGibbs}, where
$P()$ is the full probability
distribution of the
Bayesian net $\rvx$ being considered, and
$x_0$ is the starting value for $\rvx$.

\item[Number of Qubits:] This is the total
number of qubits used by the quantum circuit,
equal to $2\nb+ac$ in the notation of
Ref.\cite{TucGibbs}.

\item[Number of Elementary Operations:]
This is the number of elementary operations
in the output quantum circuit.
If there are no LOOPs, this is
the number of lines in the English File
(see Sec. \ref{sec-eng-file}), which
equals the number of lines in the
Picture File (see Sec. \ref{sec-pic-file}).
For a LOOP (assuming it is not nested
inside a larger LOOP), the
``{\tt LOOP k REPS:$N$}" and
``{\tt NEXT k}" lines are not counted,
whereas the lines between
``{\tt LOOP k REPS:$N$}" and
``{\tt NEXT k}"
are counted $N$ times (because
 {\tt REPS:$N$} indicates
 $N$ repetitions of the loop body).
Multiplexors expressed as a
single line are counted as a
single elementary operation
(unless, of course, they are inside a LOOP,
in which case they are
counted as many times as the loop body
is repeated).

\end{description}

\section{Input Files}
\label{sec-in-files}
As explained earlier,
for Quibbs to generate
quantum circuit files,
it needs to first read 3 input files:
the Parents File called {\tt parents.txt},
the States File called {\tt states.txt},
and the Probabilities File called {\tt probs.txt}.
These 3 input files
 must be placed inside the I/O folder.
Next we explain
the contents of each of these 3 input files.

\subsection{Parents File}
\begin{figure}[h]
    \begin{center}
    \includegraphics[scale=.80]{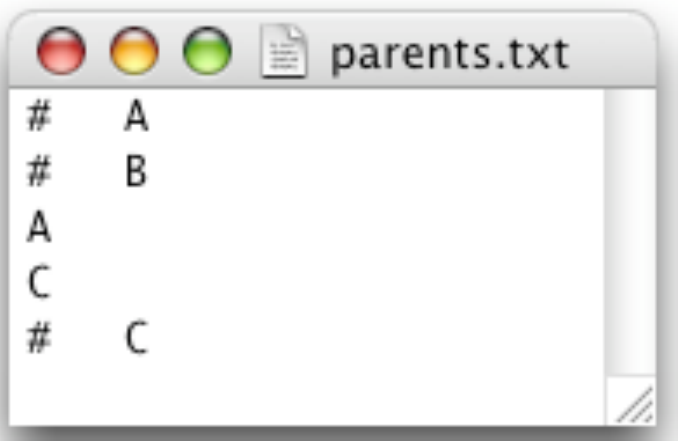}
    \caption{Parents file in the I/O folder
    ``3nodes", for a Bayesian net with graph
    $A\rarrow B \larrow C$
    }
    \label{fig-parents}
    \end{center}
\end{figure}

Fig.\ref{fig-parents}
shows the Parents File
as found in the folder ``3nodes" which is
included with the Quibbs distribution,
for a Bayesian net with graph
$A\rarrow B \larrow C$.
In this example, nodes $A$ and $C$
have no parents and node $B$ has
parents $A$ and $C$.

In general, a Parents
File must obey the
following rules:
\begin{itemize}
\item Call focus nodes the node names
immediately after a hash.
Focus nodes
in the States, Parents and Probabilities
Files must all be in the same order.
For example,
in the ``3nodes" case,
that order is $A,B,C$.

\item For each focus node, give a hash,
then the name of the focus node,
then a list of parents of
the focus node, separating all of these
with whitespace.
\end{itemize}

\subsection{States File}
\begin{figure}[h]
    \begin{center}
    \includegraphics[scale=.80]{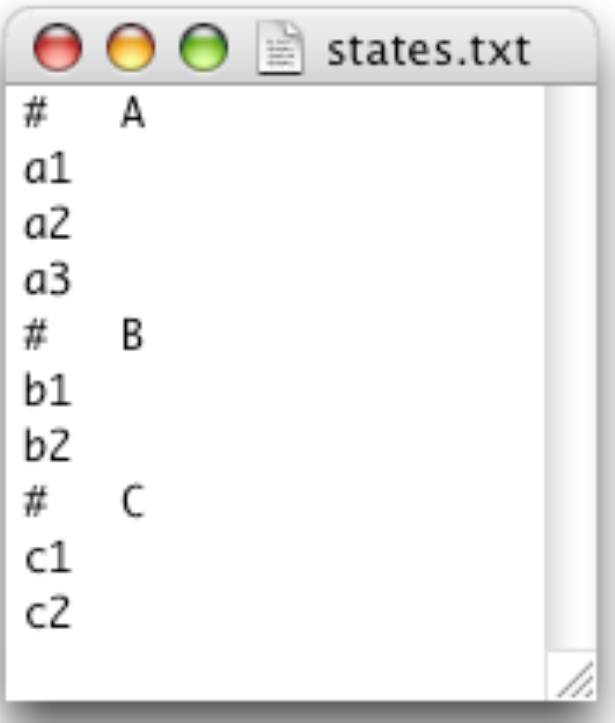}
    \caption{States file in the I/O folder
    ``3nodes", for a Bayesian net with graph
    $A\rarrow B \larrow C$
    }
    \label{fig-states}
    \end{center}
\end{figure}

Fig.
\ref{fig-states}
shows the States File
as found in the folder ``3nodes" which is
included with the Quibbs distribution,
for a Bayesian net with graph
$A\rarrow B \larrow C$.
In this example, node $A$ has 3 states
called $a1,a2$ and $a3$, node $B$ has
2 states called $b1$ and $b2$, and
node $C$ has 2 states called $c1$ and $c2$.

In general, a States
File must obey the
following rules:
\begin{itemize}
\item
Call focus nodes the node names
immediately after a hash.
Focus nodes
in the States, Parents and Probabilities
Files must all be in the same order.
For example,
in the ``3nodes" case,
that order is $A,B,C$.

\item
For each focus node, give a hash,
then the name of the focus node,
then a list of names of the states of the
focus node,
separating all of these
with whitespace.
\end{itemize}

\subsection{Probabilities File}
\label{sec-probs-file}

\begin{figure}[h]
    \begin{center}
    \includegraphics[scale=.80]{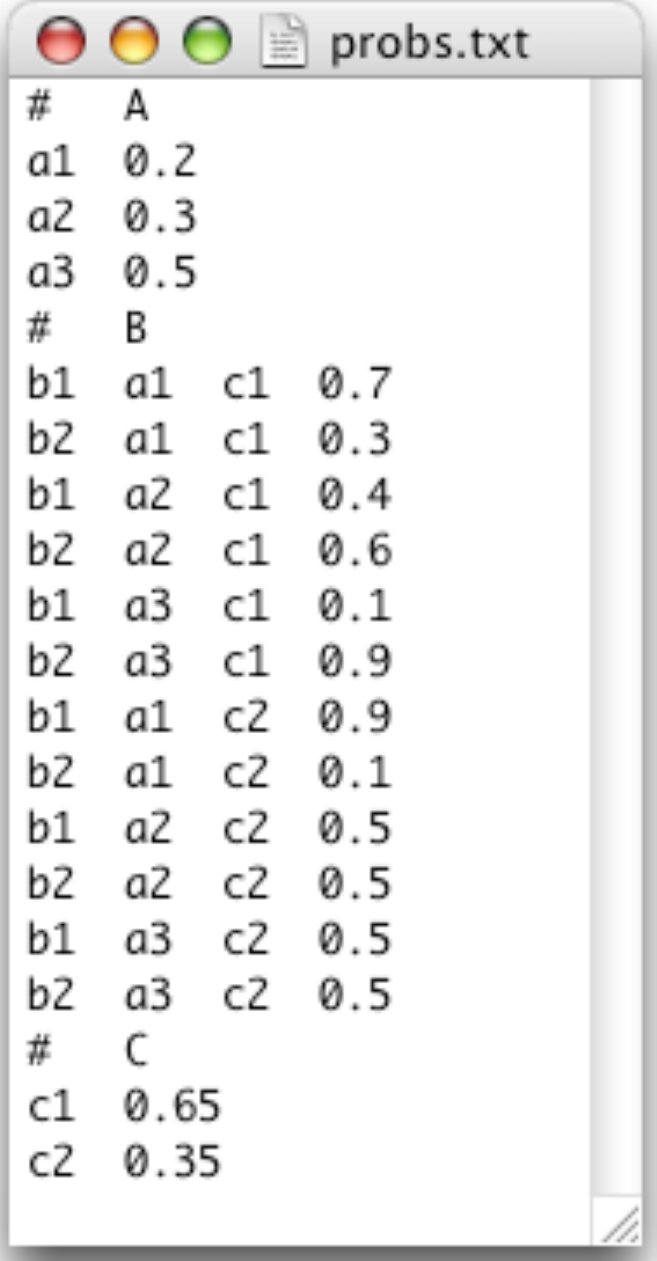}
    \caption{Probabilities File in the I/O folder
    ``3nodes", for a Bayesian net with graph
    $A\rarrow B \larrow C$
    }
    \label{fig-probs}
    \end{center}
\end{figure}
Fig.\ref{fig-probs}
shows  the Probabilities File
as found in the folder ``3nodes" which is
included with the Quibbs distribution,
for a Bayesian net with graph
$A\rarrow B \larrow C$.
In this example, $P_A(a1)=0.2$,
$P_{B|A,C}(b1|a1,c1)=0.7$, etc.

In general, a Probabilities
File must obey the
following rules:
\begin{itemize}

\item
Call focus nodes the node names
immediately after a hash.
Focus nodes
in the States, Parents and Probabilities
Files must all be in the same order.
For example,
in the ``3nodes" case,
 that order is $A,B,C$.

\item
For each focus node, give a hash,
then the name of the
focus node, then
the state of the focus node, then the states
of each parent of the focus node,
then the conditional probability of
the focus node conditioned on its parents,
separating all of these
with whitespace.

\item
The order in which the
states of the parents of the focus node
are listed must be
identical to the order in which
the parents of that
focus node are listed in the Parents File.
For example,
in the ``3nodes" case, the Parents
File gives the parents
of node $B$ as $A,C$, in that
order. Hence, in the Probabilities File,
each conditional probability for focus node $B$
is given after giving the states
of nodes $B,A,C$, in that order.

\item
A combination of node states
may be omitted, in which case Quibbs
will interpret that probability to
be zero. For example,
in ``3nodes" case,
if

$$P_{B|A,C}(b1|a2,c2)=0,$$
you could omit a line of the form

\begin{verbatim}
               b1    a2    c2   0.0
\end{verbatim}
for the focus node $B$.
\end{itemize}

Note that Quibbs can
help you to write a
Probabilities File, by
generating a template
that you can change
according to your
needs. Such templates
can be generated by
means of the {\bf Do a Pre-run}
button. See Sec.\ref{sec-pre-run}
for a detailed explanation of this.

\section{Output Files}
\label{sec-out-files}

As explained earlier, when you press the
{\bf Write Q. Circuit Files} button,
Quibbs writes 3 output files
within the I/O folder:
a Log File called {\tt quibbs\_log.txt},
an English File called {\tt quibbs\_eng.txt},
and a Picture File called {\tt quibbs\_pic.txt}.
Next we explain
the contents of each of these 3 output files.
We also explain
the contents of the various
files generated when you
press the {\bf Do a Pre-run} button.

\subsection{Log File}
\begin{figure}[h]
    \begin{center}
    \includegraphics[scale=.70]{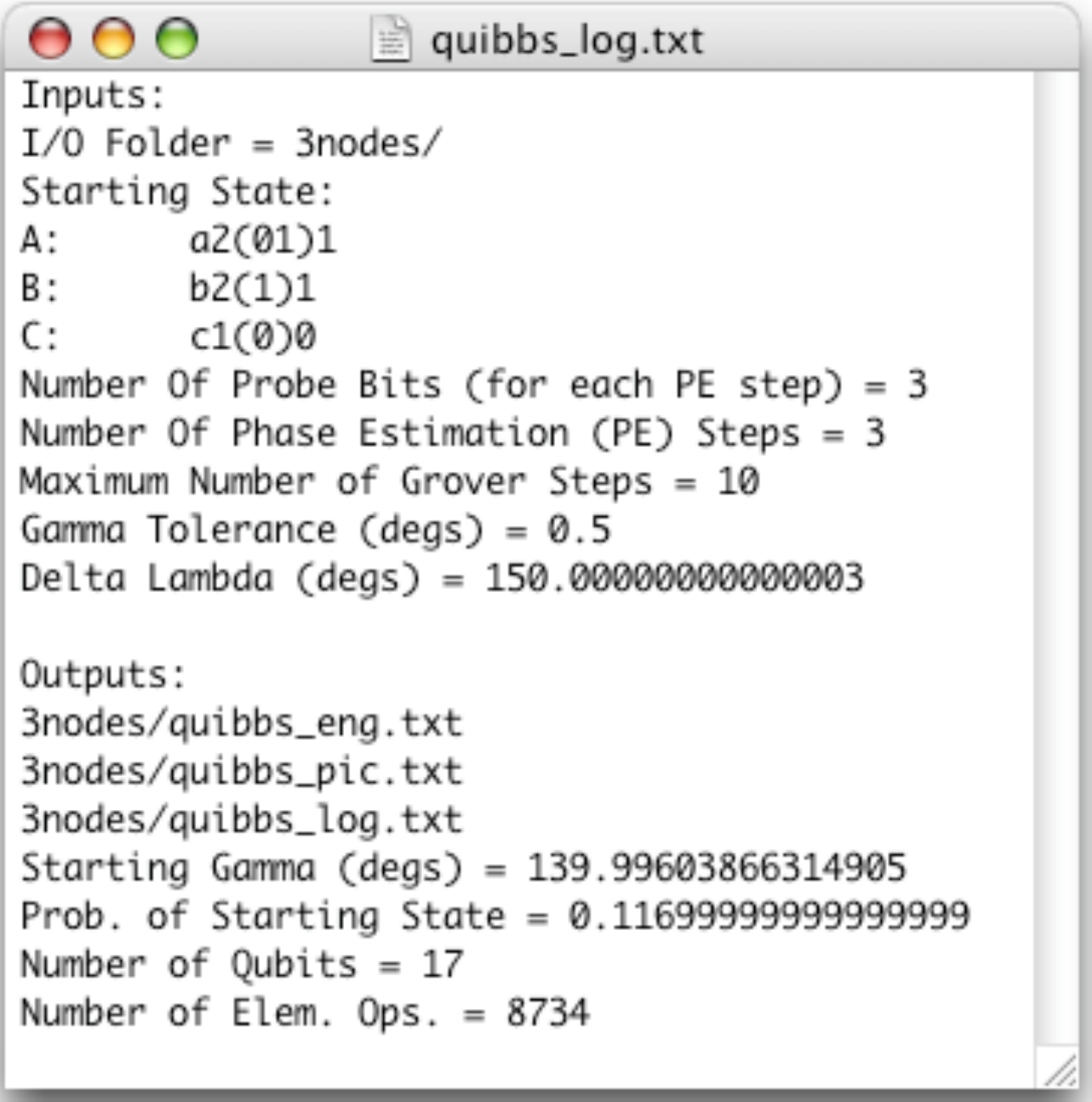}
    \caption{
    Log File generated by Quibbs
     using input files from the
    ``3nodes" I/O folder.}
    \label{fig-quibbs-log}
    \end{center}
\end{figure}

Fig.\ref{fig-quibbs-log}
is an example a Log File.
This example was
generated by
Quibbs using the input files from the
``3nodes" I/O folder.
A Log File
records all the information
found in the
{\bf Control Panel}.

\subsection{English File}\label{sec-eng-file}
\begin{figure}[h]
    \begin{center}
    \includegraphics[scale=.70]{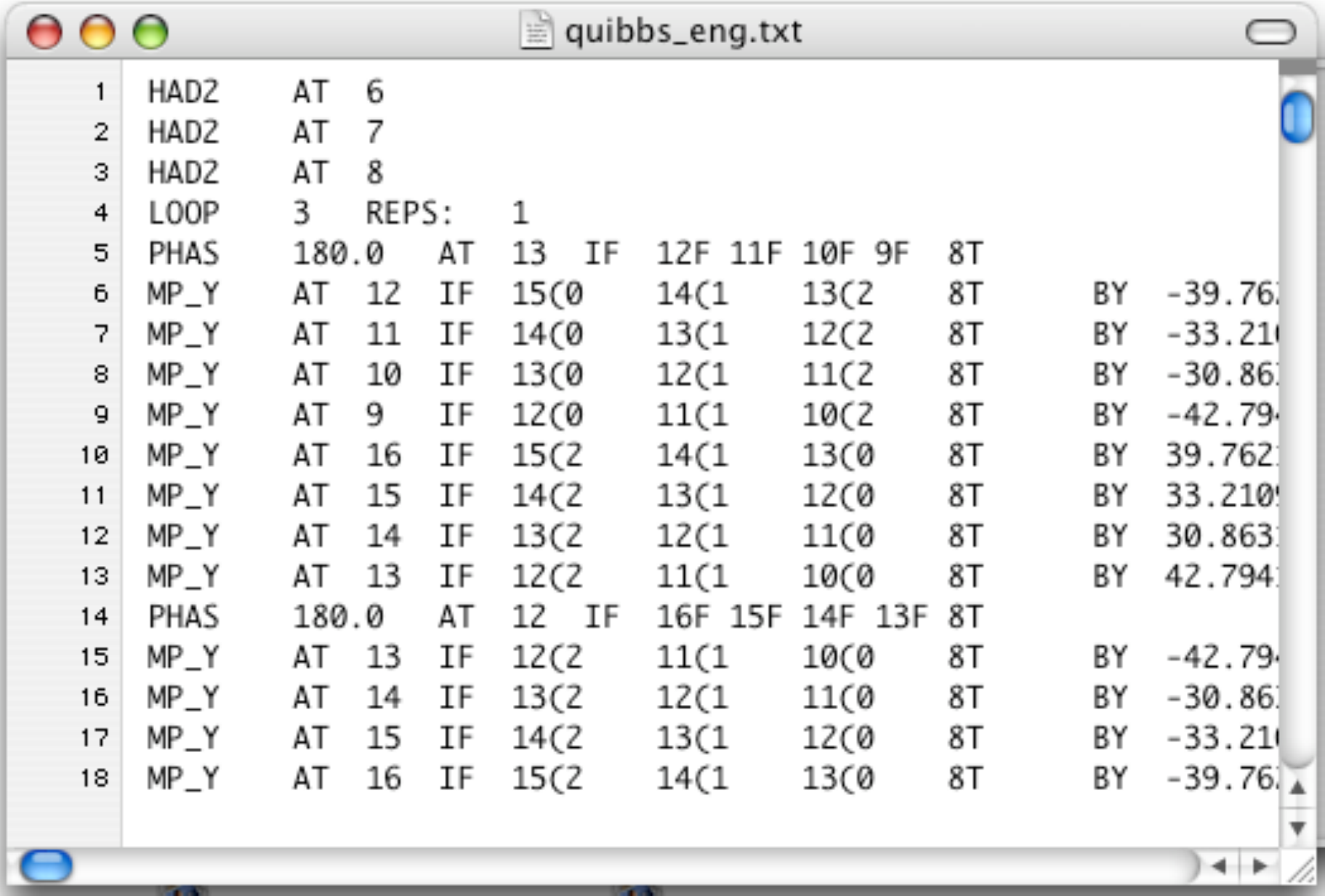}
    \caption{
    English File generated by
    Quibbs (with the {\bf Omit V Gates} feature OFF)
    in the same run as the
     Log File of Fig.\ref{fig-quibbs-log}, and
    using the  input files from the
    ``3nodes" I/O folder.
     Bottom of file is not visible.
     Right hand side of file is not visible.
    }
    \label{fig-quibbs-eng}
    \end{center}
\end{figure}

\begin{figure}[h]
    \begin{center}
    \includegraphics[scale=.70]{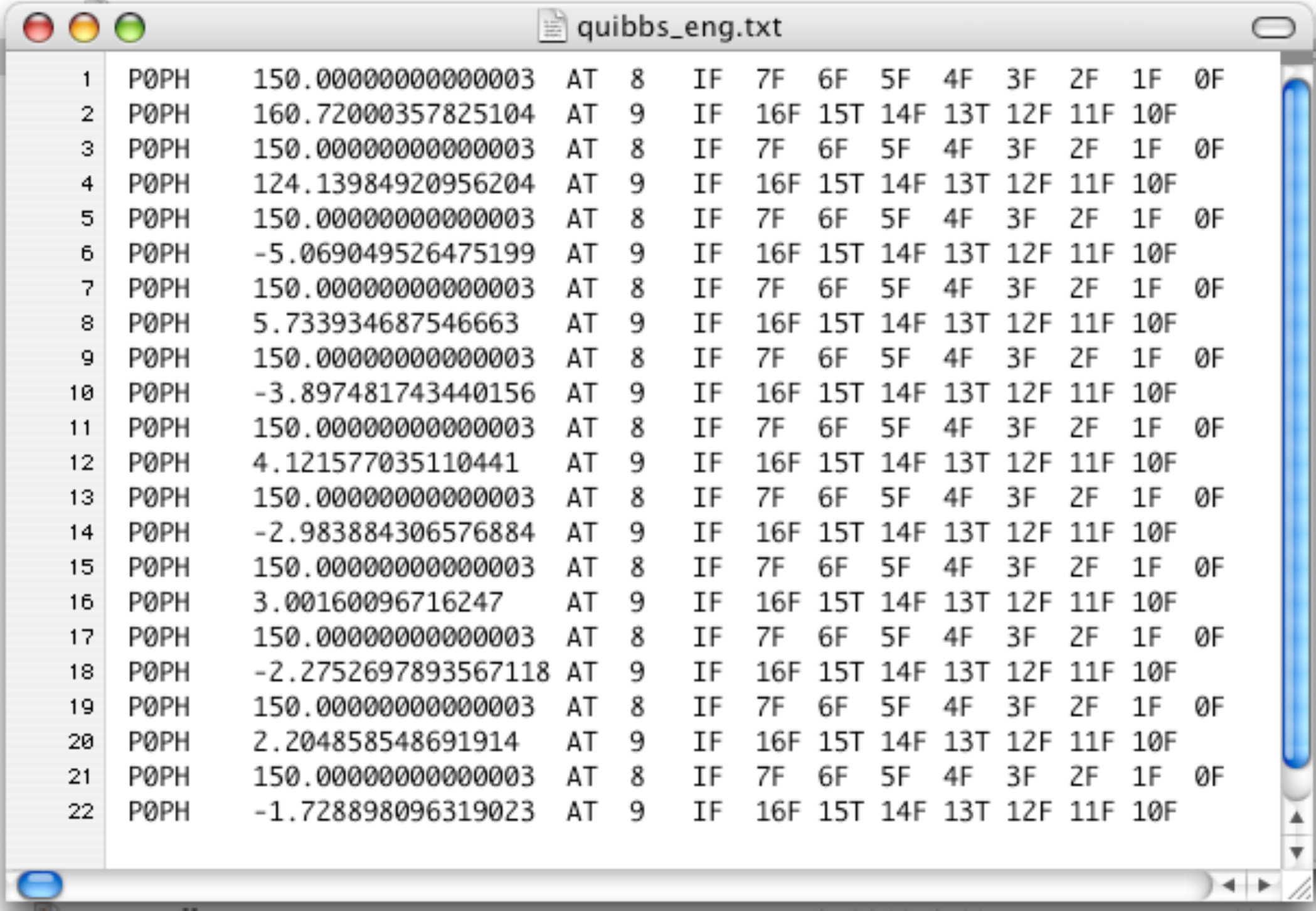}
    \caption{
    English File generated by
    Quibbs (with the {\bf Omit V Gates} feature ON)
    in the same run as the
     Log File of Fig.\ref{fig-quibbs-log}, and
    using the  input files from the
    ``3nodes" I/O folder.
     Bottom of file is visible.
     Right hand side of file is visible.
    }
    \label{fig-quibbsOmit-eng}
    \end{center}
\end{figure}

Fig.\ref{fig-quibbs-eng}
(respectively, Fig.\ref{fig-quibbsOmit-eng})
is an example of an English File.
This example was
generated by
Quibbs,
with the {\bf Omit V Gates} feature
OFF (respectively, ON),
 in the same run as the
Log File of Fig.\ref{fig-quibbs-log},
and using the input files from the
``3nodes" I/O folder.
An English File
completely specifies the output SEO.
It does so ``in English", thus its name.
Each line represents one elementary operation,
and time increases as we move downwards.

In general, an English File obeys
the following rules:

\begin{itemize}
\item Time grows as we move down the file.

\item Each row
corresponds to one elementary operation.
Each row starts with 4 letters that indicate
the type of elementary operation.

\item For a one-bit operation
acting on a ``target bit" $\alpha$,
the target bit $\alpha$ is
given after the word {\tt AT}.

\item If the one-bit operation is controlled, then
the controls are indicated after the word {\tt IF}.
{\tt T} and {\tt F} stand for
true and false, respectively.
{\tt $\alpha$T} stands for
a control $P_1(\alpha)=n(\alpha)$ at bit $\alpha$.
{\tt $\alpha$F} stands for
a control $P_0(\alpha)=\nbar(\alpha)$ at bit $\alpha$.

\item ``{\tt LOOP k REPS:$N$}" and ``{\tt NEXT k}"
mark the beginning and end of $N$
repetitions. {\tt k} labels the loop. {\tt k} also
equals the line-count number in the English file
(first line is 0)
of the line
``{\tt LOOP k REPS:$N$}".

\item {\tt SWAP
$\alpha$ $\beta$}
stands for the swap(i.e., exchange) operator
$E(\alpha, \beta)$
that swaps
bits $\alpha$ and $\beta$.

\item {\tt PHAS }$\theta^{degs}$ stands for
a phase factor $e^{i \theta^{degs} \frac{\pi}{180}}$.

\item
{\tt P0PH }$\theta^{degs}$ stands for
the one-bit gate
 $e^{i P_0 \theta^{degs} \frac{\pi}{180}}$ (note
 $P_0=\nbar$).
{\tt P1PH }$\theta^{degs}$ stands for
  the one-bit gate
 $e^{i P_1 \theta^{degs} \frac{\pi}{180}}$
 (note $P_1=n$).
 Target bit follows the word {\tt AT}.

\item {\tt SIGX}, {\tt SIGY},
{\tt SIGZ}, {\tt HAD2}
stand for
the Pauli matrices $\sigx, \sigy, \sigz$
and the one-bit Hadamard matrix $H$,
respectively.
Target bit follows the word {\tt AT}.

\item {\tt ROTX}, {\tt ROTY},
{\tt ROTZ}, {\tt ROTN}
stand for one-bit
rotations
with rotation axes in the
directions: $x$, $y$, $z$, and
an arbitrary direction $n$, respectively.
Rotation angles (in degrees) follow
the words {\tt ROTX}, {\tt ROTY},
{\tt ROTZ}, {\tt ROTN}.
Target bit follows the word {\tt AT}.

\item
{\tt MP\_Y} stands for a multiplexor
which performs a one-bit rotation of
a target bit about the
$y$ axis. Target bit follows the word {\tt AT}.
Rotation angles (in degrees) follow
the word {\tt BY}. Multiplexor controls
are specified by $\alpha(k$, where
integer $\alpha$ is the bit position
and integer $k$ is the control's name.

\end{itemize}

Here is a list of examples
showing how to translate the mathematical
notation used in Ref.\cite{notationNAND}
into the English File language:
\begin{center}
\begin{tabular}{|l|l|}
\hline
Mathematical language & English File language\\
\hline
\hline
Loop named 5 with 2 repetitions &
{\tt  LOOP 5 REPS: 2}\\
\hline
Next iteration of loop named 5&
{\tt  NEXT 5}\\
\hline
$E(1,0)^{\nbar(3)n(2)}$ &
{\tt SWAP  1  0  IF  3F  2T}\\
\hline
$e^{i 42.7 \frac{\pi}{180} \nbar(3)n(2)}$ &
{\tt  PHAS 42.7 IF  3F  2T}\\
\hline
$e^{i 42.7 \frac{\pi}{180} \nbar(3)n(2)}$ &
{\tt  P0PH 42.7 AT  3 IF 2T}\\
\hline
$e^{i 42.7 \frac{\pi}{180} n(3)n(2)}$ &
{\tt  P1PH 42.7 AT  3 IF 2T}\\
\hline
$\sigx(1)^{\nbar(3)n(2)}$ &
{\tt  SIGX  AT  1  IF  3F  2T}\\
\hline
$\sigy(1)^{\nbar(3)n(2)}$ &
{\tt  SIGY  AT  1  IF  3F  2T}\\
\hline
$\sigz(1)^{\nbar(3)n(2)}$ &
{\tt  SIGZ  AT  1  IF  3F  2T}\\
\hline
$H(1)^{\nbar(3)n(2)}$ &
{\tt  HAD2  AT  1  IF  3F  2T}\\
\hline
$(e^{i \frac{\pi}{180} 23.7 \sigx(1)})^{\nbar(3)n(2)}$ &
{\tt  ROTX  23.7  AT  1  IF  3F  2T}\\
\hline
$(e^{i  \frac{\pi}{180} 23.7 \sigy(1)})^{\nbar(3)n(2)}$ &
{\tt  ROTY  23.7  AT  1  IF  3F  2T} \\
\hline
$(e^{i  \frac{\pi}{180} 23.7 \sigz(1)})^{\nbar(3)n(2)}$ &
{\tt  ROTZ  23.7  AT  1  IF  3F  2T}\\
\hline
$(e^{
i  \frac{\pi}{180}
[30\sigx(1)+ 40\sigy(1) + 11 \sigz(1)]
 })^{\nbar(3)n(2)}$ &
{\tt  ROTN  30.0 40.0 11.0  AT  1  IF  3F  2T}\\
\hline
$[e^{i\sum_{b1,b0}\theta_{b_1b_0}\sigy(3)P_{b_1b_0}(2,1)}]^{n(0)}$
&
{\tt MP\_Y  AT  3 IF 2(1 1(0 0T BY 30.0 10.5 11.0 83.1}
\\
where $\left\{\begin{array}{l}
\theta_{00}=30.0(\frac{\pi}{180})
\\
\theta_{01}=10.5(\frac{\pi}{180})
\\
\theta_{10}=11.0(\frac{\pi}{180})
\\
\theta_{11}=83.1(\frac{\pi}{180})
\end{array}\right.$
&\;
\\
\hline
\end{tabular}
\end{center}

\subsection{ASCII Picture File}\label{sec-pic-file}

\begin{figure}[h]
    \begin{center}
    \includegraphics[scale=.70]{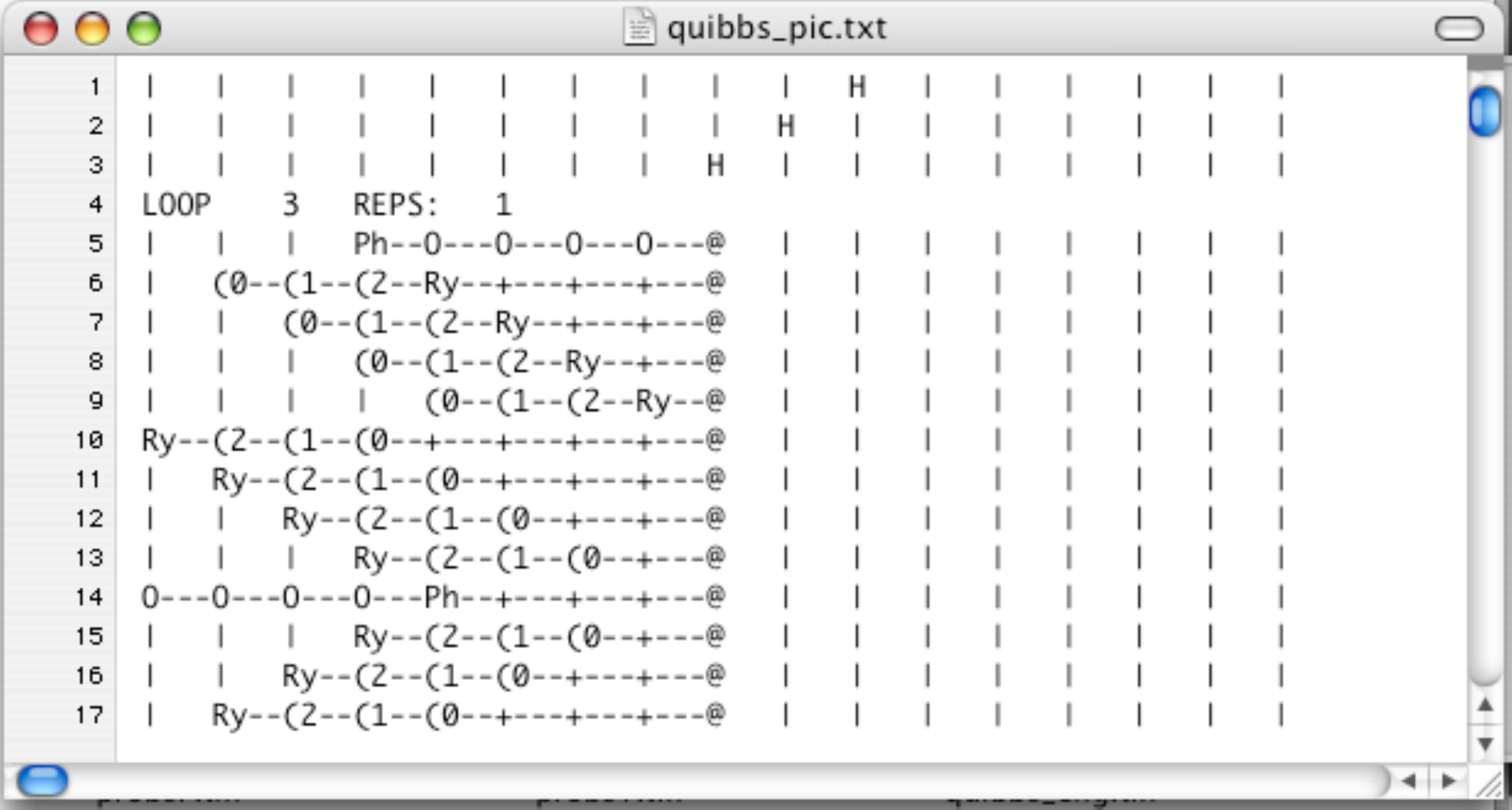}
    \caption{
    Picture File generated by
    Quibbs (with the {\bf Omit V Gates} feature OFF)
     in the same run as the
     Log File of Fig.\ref{fig-quibbs-log}, and
     using input files from the
    ``3nodes" I/O folder.
     Bottom of file is not visible.
    }
    \label{fig-quibbs-pic}
    \end{center}
\end{figure}

\begin{figure}[h]
    \begin{center}
    \includegraphics[scale=.70]{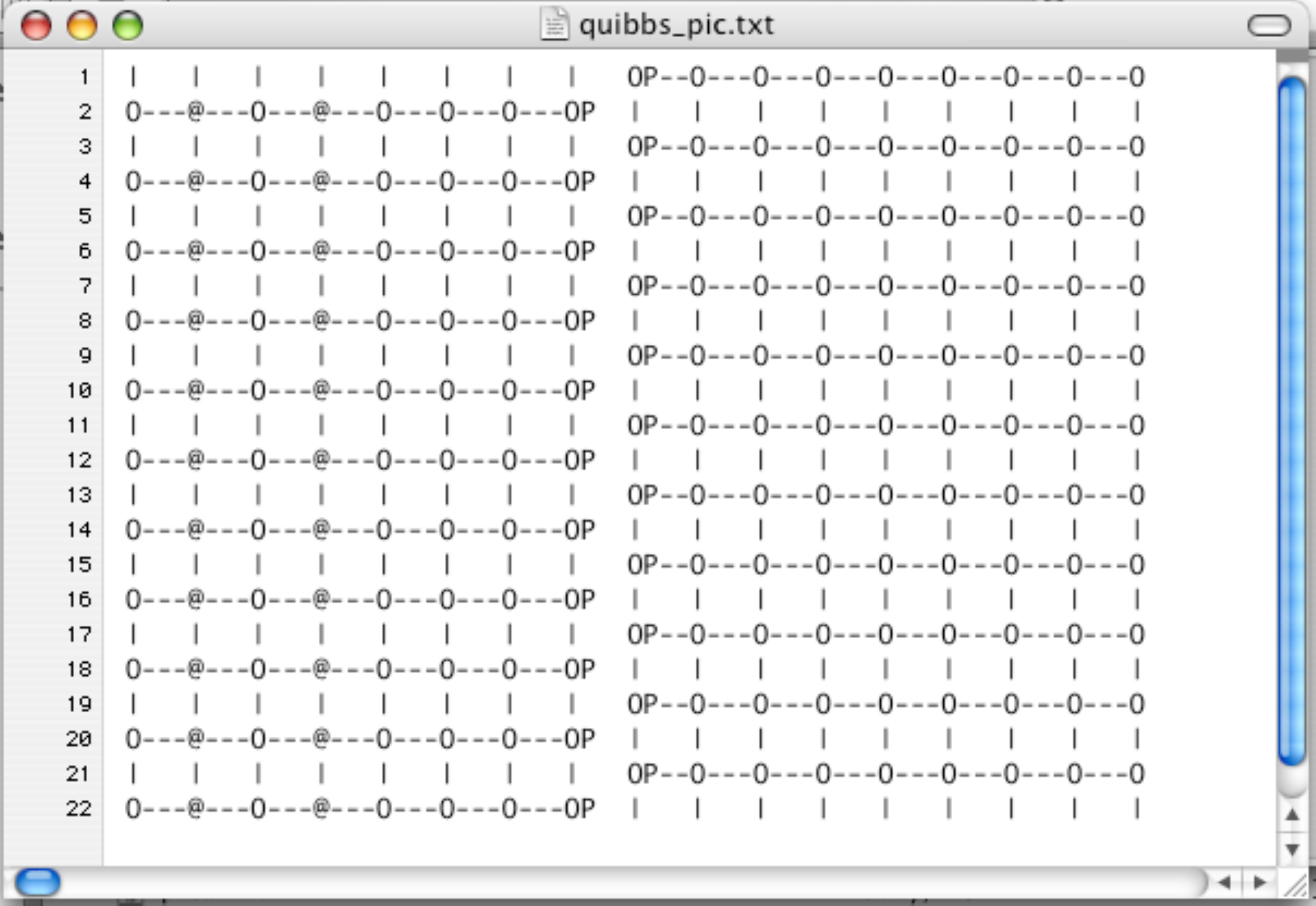}
    \caption{
    Picture File generated by
    Quibbs (with the {\bf Omit V Gates} feature ON)
     in the same run as the
     Log File of Fig.\ref{fig-quibbs-log}, and
     using input files from the
    ``3nodes" I/O folder.
     Bottom of file is visible.
    }
    \label{fig-quibbsOmit-pic}
    \end{center}
\end{figure}

Fig.\ref{fig-quibbs-pic}
(respectively, Fig.\ref{fig-quibbsOmit-pic})
is an example of a Picture File.
This example was
generated by
Quibbs,
with the {\bf Omit V Gates} feature
OFF (respectively, ON),
in the same run as the
Log File of Fig.\ref{fig-quibbs-log},
and using the input files from the
``3nodes" I/O folder.
A Picture File
partially specifies the output SEO.
It gives an ASCII picture of
the quantum circuit.
Each line represents one elementary operation,
and time increases as we move downwards.
There is a one-to-one onto correspondence
between the rows of the English
and Picture Files.

In general, a Picture File obeys
the following rules:

\begin{itemize}
\item Time grows as we move down the file.

\item Each row
corresponds to one elementary operation.
Columns $1, 5, 9, 13, \ldots$ represent
qubits (or, qubit positions). We define the
rightmost qubit  as 0. The qubit
immediately to
the left of the rightmost qubit
is 1, etc.
For a one-bit operator
acting on a ``target bit" $\alpha$,
one places a symbol
of the operator at bit position
$\alpha$.

\item {\tt |} represents a ``qubit wordline"
connecting the same qubit at
two consecutive times.

\item {\tt -}represents a wire connecting different
qubits at the same time.

\item{\tt +} represents both {\tt |} and {\tt -}.

\item If the one-bit operation is controlled, then
the controls are indicated
as follows.
{\tt @} at bit position $\alpha$ stands for
a control $n(\alpha)=P_1(\alpha)$.
{\tt 0} at bit position $\alpha$ stands for
a control $\nbar(\alpha)=P_0(\alpha)$.

\item ``{\tt LOOP k REPS:$N$}" and ``{\tt NEXT k}"
mark the beginning and end of $N$
repetitions. {\tt k} labels the loop. {\tt k} also
equals the line-count number in the Picture File
(first line is 0)
of the line
``{\tt LOOP k REPS:$N$}" .

\item The swap(i.e., exchange) operator
$E(\alpha, \beta)$
is represented by putting
arrow heads {\tt <} and {\tt >} at
bit positions $\alpha$ and $\beta$.

\item A
phase factor $e^{i\theta}$ for
$\theta\in \RR$ is represented by
placing {\tt Ph} at any bit position
which does not already hold a control.

\item The one-bit gate
$e^{i P_0(\alpha)\theta}$ (note
$P_0(\alpha)=\nbar(\alpha)$) for $\theta\in \RR$
is represented by putting {\tt OP}
at bit position $\alpha$.

\item The one-bit gate
$e^{i P_1(\alpha)\theta}$
(note
$P_1(\alpha)=n(\alpha)$) for $\theta\in \RR$
is represented by putting {\tt @P}
at bit position $\alpha$.

\item One-bit operations
 $\sigx(\alpha)$,
 $\sigy(\alpha)$,
 $\sigz(\alpha)$
and $H(\alpha)$
are represented by placing the letters
{\tt X,Y,Z, H}, respectively,
at bit position $\alpha$.

\item
One-bit rotations
acting on bit $\alpha$,
in the
$x,y,z,n$ directions,
are represented by placing
{\tt Rx,Ry,Rz, R}, respectively,
at bit position $\alpha$.

\item
A multiplexor that rotates
a bit $\tau$ about the $y$ axis
is represented by placing
{\tt Ry}
at bit position $\tau$.
A multiplexor control at bit position $\alpha$
and
named by the integer $k$
is represented by placing
$(k$ at bit position $\alpha$.
\end{itemize}

Here is a list of examples
showing how to translate the mathematical
notation used in Ref.\cite{notationNAND}
into the Picture File language:

\begin{tabular}{|l|l|}
\hline
Mathematical language & Picture File language\\
\hline
\hline
Loop named 5 with 2 repetitions &
{\tt  LOOP 5 REPS:2}\\
\hline
Next iteration of loop named 5&
{\tt  NEXT 5}\\
\hline
$E(1,0)^{\nbar(3)n(2)}$& {\tt 0---@---<--->} \\
\hline
$e^{i 42.7 \frac{\pi}{180} \nbar(3)n(2)}$ &
{\tt 0---@---+--Ph}\\
\hline
$e^{i 42.7 \frac{\pi}{180} \nbar(3)n(2)}$ &
{\tt 0P--@\ \ \ |\ \ \ |}\\
\hline
$e^{i 42.7 \frac{\pi}{180} n(3)n(2)}$ &
{\tt @P--@\ \ \ |\ \ \ |}\\
\hline

 $\sigx(1)^{\nbar(3)n(2)}$& {\tt 0---@---X\ \ \ |} \\
\hline
 $\sigy(1)^{\nbar(3)n(2)}$& {\tt 0---@---Y\ \ \ |} \\
\hline
 $\sigz(1)^{\nbar(3)n(2)}$& {\tt 0---@---Z\ \ \ |} \\
\hline
$H(1)^{\nbar(3)n(2)}$& {\tt 0---@---H\ \ \ |} \\
\hline
$(e^{i \frac{\pi}{180} 23.7 \sigx(1)})^{\nbar(3)n(2)}$&
{\tt 0---@---Rx\ \ |} \\
\hline
$(e^{i  \frac{\pi}{180} 23.7 \sigy(1)})^{\nbar(3)n(2)}$&
{\tt 0---@---Ry\ \ |} \\
\hline
$(e^{i  \frac{\pi}{180} 23.7 \sigz(1)})^{\nbar(3)n(2)}$&
{\tt 0---@---Rz\ \ |} \\
\hline
$(e^{
i  \frac{\pi}{180}
[30\sigx(1)+ 40\sigy(1) + 11 \sigz(1)]
})^{\nbar(3)n(2)}$&
{\tt 0---@---R\ \ \ |} \\
\hline
$[e^{i\sum_{b1,b0}\theta_{b_1b_0}\sigy(3)P_{b_1b_0}(2,1)}]^{n(0)}$
&
{\tt |\ \ \ Ry--(1--(0--@}
\\
where $\left\{\begin{array}{l}
\theta_{00}=30.0(\frac{\pi}{180})
\\
\theta_{01}=10.5(\frac{\pi}{180})
\\
\theta_{10}=11.0(\frac{\pi}{180})
\\
\theta_{11}=83.1(\frac{\pi}{180})
\end{array}\right.$
&\;
\\
\hline
\end{tabular}

\subsection{Output Files From a Pre-run}
\label{sec-pre-run}
When you
press the {\bf Do a Pre-run} button,
Quibbs writes 4 output files
within the I/O folder:
two Uniform Probabilities Files
called
{\tt probsF.txt} and
{\tt probsT.txt}, a Blankets File
called
{\tt blankets.txt}, and a Nits File called
{\tt nits.txt}.
Next we explain
the contents of each of these 4 output files.

\subsubsection{Uniform Probabilities Files}
\begin{figure}[h]
    \begin{center}
    \includegraphics[
    scale=.80
    ]{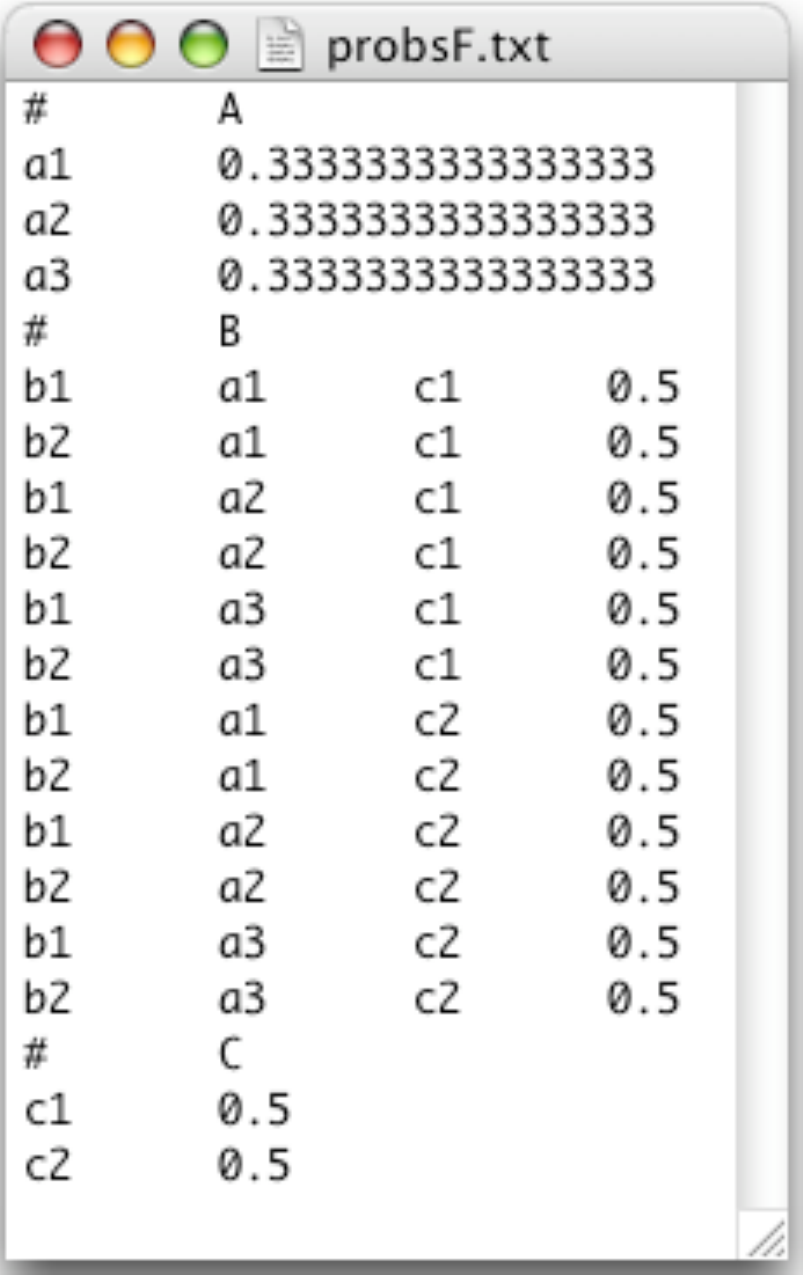}
    \caption{
    Uniform Probabilities File
    (of type {\tt probsF.txt})
    generated using input files from the
    ``3nodes" I/O folder.
    }
    \label{fig-probsF}
    \end{center}
\end{figure}

\begin{figure}[h]
    \begin{center}
    \includegraphics[
    scale=.80
    ]{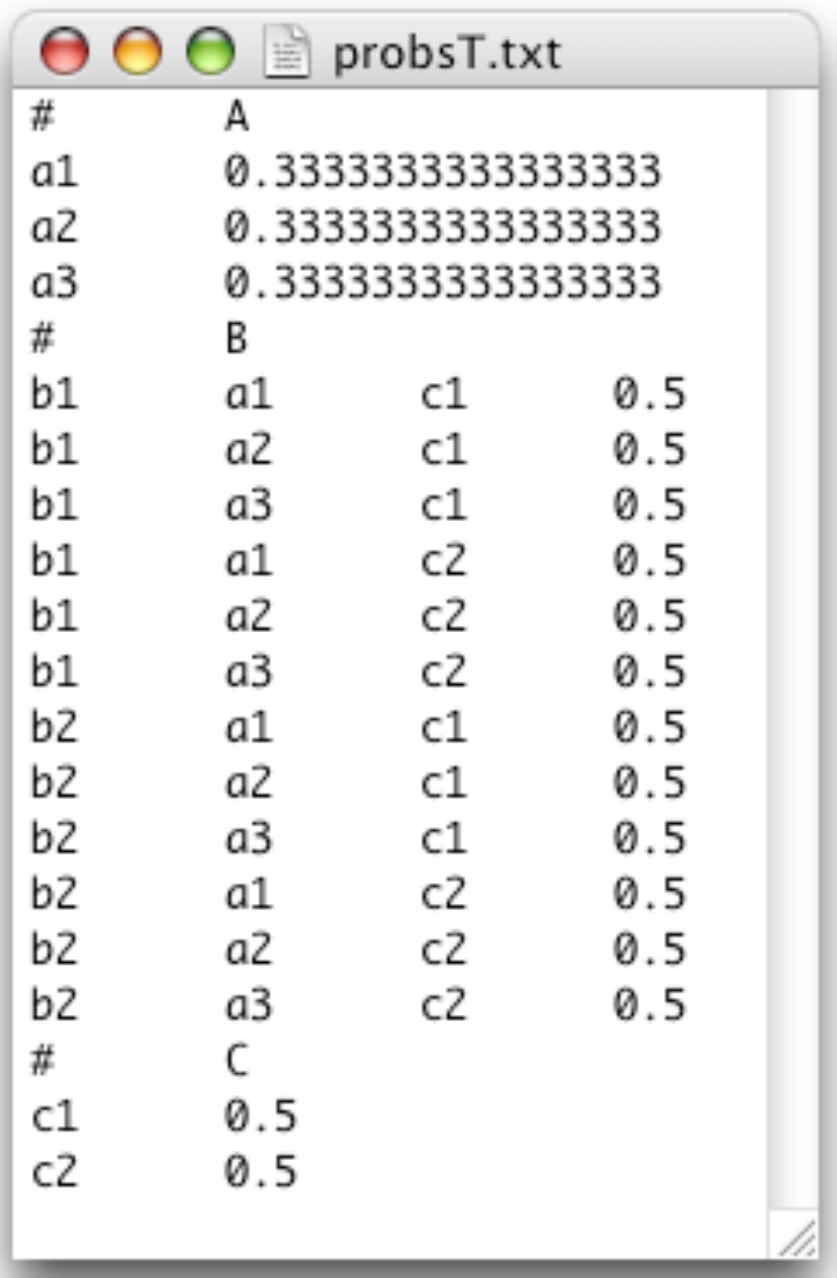}
    \caption{Uniform Probabilities File
    (of type {\tt probsT.txt})
    generated using input files from the
    ``3nodes" I/O folder.
    }
    \label{fig-probsT}
    \end{center}
\end{figure}

Figs.\ref{fig-probsF}
and \ref{fig-probsT}
are both examples of a
Uniform Probabilities File.
Both files were
generated by
Quibbs using the input files from the
``3nodes" I/O folder.
A Uniform Probabilities File
is simply a Probability File,
as defined in Sec.\ref{sec-probs-file},
but of a specific kind
that assigns uniform
values to all conditional
probabilities of the Bayesian net
specified by the Parents File and
States File in the I/O Folder.
Generating a
Uniform
Probabilities File
(either {\tt probsF.txt} or {\tt probsT.txt})
does not require a {\tt probs.txt}
file. One can use a
Uniform Probabilities File
as a template
for a {\tt probs.txt} file.
Just cut and paste the contents of a
Uniform Probabilities File into a new
file called {\tt probs.txt}
and modify its probabilities according
to your needs.
Note that the only
difference between a {\tt probsF.txt}
and a {\tt probsT.txt} file
is that the first (respectively, second) of these
varies the states of a focus
node before (respectively, after)
varying the states of its parents.

\subsubsection{Blankets File}
\begin{figure}[h]
    \begin{center}
    \includegraphics[
    scale=.80
    ]{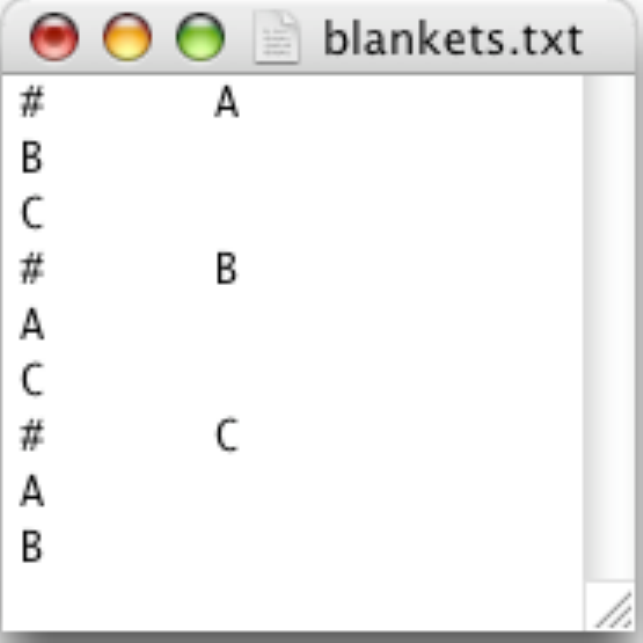}
    \caption{Blankets File
    generated using input files from the
    ``3nodes" I/O folder.
    }
    \label{fig-blankets}
    \end{center}
\end{figure}

Fig.\ref{fig-blankets}
is an example of a
Blankets File.
This example was
generated by
Quibbs using the input files from the
``3nodes" I/O folder.

The Markov blanket
$MB(i)$ for a node $\rvx_i$
of the classical Bayesian network $\rvx$ is defined
so that
(see section entitled
``Notation and Preliminaries"
in  Ref.\cite{TucMetHas})

\beq
P(x_i|x_\noti)=P(x_i|x_{MB(i)})
\;.
\label{eq-mb}
\eeq
It can be shown that
the Markov blanket of a
focus node equals the union
of:
\begin{itemize}
\item the parents
of the focus node
\item
the
children of the focus node
\item
the parents of each children
of the focus node (but
excluding the focus node itself)
\end{itemize}

A Blankets File gives the Markov
blanket for each node of a Bayesian net.
In the example of
Fig.\ref{fig-blankets},
node $A$ has Markov blanket
$\{B,C\}$, etc.

In general, a Blankets File obeys the
following rules:
\begin{itemize}
\item Call focus nodes the node names
immediately after a hash.

\item For each focus node, Quibbs writes a hash,
then the name of the focus node,
then a list of the nodes which form the
Markov blanket
of
the focus node.
\end{itemize}

\subsubsection{Nits File}

\begin{figure}[h]
    \begin{center}
    \includegraphics[
    scale=.80
    ]{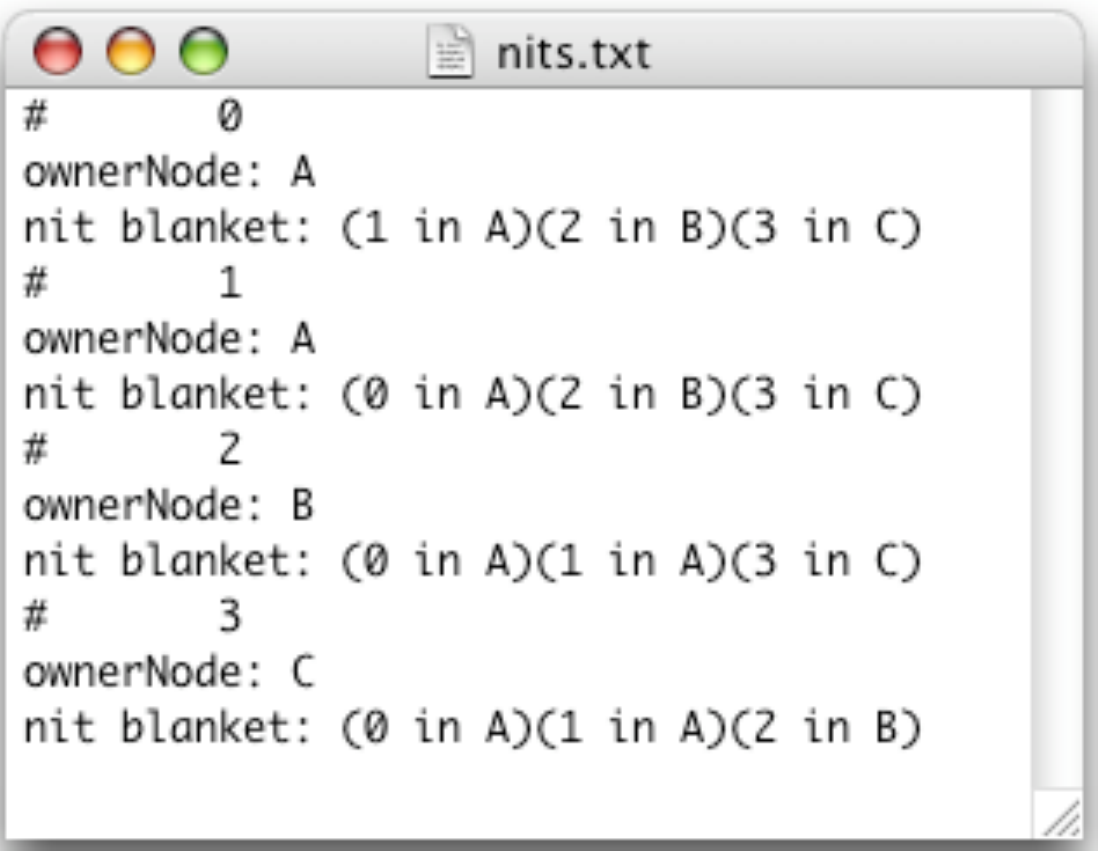}
    \caption{Nits File
    generated using input files from the
    ``3nodes" I/O folder.
    }
    \label{fig-nits}
    \end{center}
\end{figure}

Fig.\ref{fig-nits}
is an example of a
Nits File.
This example was
generated by
Quibbs using the input files from the
``3nodes" I/O folder.

The word ``nit" is a contraction
of the words ``node" and ``qubit".
Quibbs assigns to each node
(of the Bayesian net being
consider) its own private set of nits.
We explained in Sec.\ref{sec-control-panel}
how
Quibbs assigns a decimal and a binary name
to each state of a node.
The binary name of a state gives the
states of the nits. For example,
suppose node $A$ has 3 states: $a1(00)0$,
$a1(01)1$ and $a1(10)2$. Then
node $A$ is assigned two private nits,
call them $nit0$ and $nit1$.
When node $A$ is in state
$(b1,b0)$, where $b1$ and $b0$
are either 0 or 1, then
$nit1$ is in state $b1$
and $nit0$ is in state $b0$.

Actually, Quibbs doesn't
give nits an ``english" name like
$nit0$ and $nit1$. It just calls them
by integers.
Fig.\ref{fig-nits} informs us
that the ``3nodes" Bayesian net
has 4 nits called 0,1,2,3.
Nits 0 and 1 are both owned by
node $A$ (which has 3 states $a1,a2,a3$).
Nit 2 is owned by node $B$ (which has 2 states $b1,b2$).
Nit 3 is owned by node $C$ (which has 2 states $c1,c2$).

The original Bayesian net
with the original nodes implies
a new, finer Bayesian net whose nodes
are the nits themselves. Just
like one can define a Markov blanket
for each node of the original Bayesian
net, one can define a Markov blanket (
equal to a
particular set
of nits) for each nit.
Fig.\ref{fig-nits} informs us
that for the ``3nodes" Bayesian net,
nit 0 has a nit blanket $\{1,2,3\}$,
etc.

In general, a Nits File obeys the
following rules:
\begin{itemize}
\item Call focus nit the number (
a sort of nit name)
immediately after a hash.

\item For each focus nit, Quibbs writes
the words ``owner node" followed by the name
of the owner node of the focus nit.

\item For each focus nit,
Quibbs writes
the words ``blanket nit" followed by
the Markov blanket of nits
for the focus nit.

\end{itemize}

Why do we care about
node blankets and nit blankets at all?
Quibbs uses a method,
discussed in Ref.\cite{TucGibbs},
of representing
Szegedy operators
using quantum multiplexors.
Quibbs uses nit blankets
to simplify its Szegedy representations
by eliminating certain unnecessary
controls in their multiplexors.

\end{document}